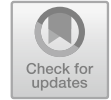

# Symmetry Reduction for the Local Mu-Calculus

Kedar S. Namjoshi[1](✉) and Richard J. Trefler[2](✉)

[1] Bell Labs, Nokia, Murray Hill, USA
kedar.namjoshi@nokia-bell-labs.com
[2] University of Waterloo, Waterloo, Canada
trefler@uwaterloo.ca

**Abstract.** Model checking large networks of processes is challenging due to state explosion. In many cases, individual processes are isomorphic, but there is insufficient global symmetry to simplify model checking. This work considers the verification of local properties, those defined over the neighborhood of a process. Considerably generalizing earlier results on invariance, it is shown that all local mu-calculus properties, including safety and liveness properties, are preserved by neighborhood symmetries. Hence, it suffices to check them locally over a set of representative process neighborhoods. In general, local verification approximates verification over the global state space; however, if process interactions are outward-facing, the relationship is shown to be exact. For many network topologies, even those with little global symmetry, analysis with representatives provides a significant, even exponential, reduction in the cost of verification. Moreover, it is shown that for network families generated from building-block patterns, neighborhood symmetries are easily determined, and verification over the entire family reduces to verification over a finite set of representative process neighborhoods.

## 1 Introduction

Networks of communicating processes are a model for distributed systems, cloud computing environments, routing protocols, many-core hardware processors, and other such systems. Often, networks are described parametrically, that is, a process template is instantiated at each node of a network graph. The expectation then is that basic correctness properties should hold regardless of the size and the shape of the network.

Model checkers can determine, fully automatically, whether a fixed instance of a process network satisfies a correctness property. However, model checking suffers from exponential state explosion as the size of the analyzed network increases. Thus, one may aim for parameteric analysis of a network family, "in one fell swoop"; however, the parametric model checking problem (PMCP) is undecidable in general [2]. Limiting to *compositional* proofs makes parametrized verification more tractable; as shown in [20], the PCMCP (Parameterized Compositional Model Checking problem) can be solved efficiently for standard network families (rings, tori, wrap-around mesh, etc.) where the PMCP is undecidable even for invariance properties.

In this work, we generalize these results considerably, from invariance to mu-calculus properties. We formulate a local version of the mu-calculus to describe behaviors of a single process and its immediate neighborhood. The logic allows specification of safety and liveness properties, each property being limited to assertions over a fixed process neighborhood – e.g., "A hungry philosopher eventually acquires all adjacent forks". The goal of this work is a method to prove such properties for all processes in a network and, moreover, to prove properties parametrically, i.e., for all networks in a family.

Our analysis is based on a grouping of processes by local symmetry, where "balanced" processes have (recursively) similar neighborhoods [17,18,20]. Such symmetries are common in parametric network structures, for example [18,19], *c.f.* [17,20]. We establish that the local state spaces of balanced processes are sufficiently bisimilar that they satisfy the same local mu-calculus properties. It is, therefore, enough to model-check a representative process from each balance class, while paying particular attention to 'interference' transitions from neighboring processes.

We show that any *universal* local mu-calculus property established locally also holds on the global state space. Thus, a universal property can be established globally for all processes by checking it on the local state spaces of a few representatives.

Many communication protocols are designed in such a way that a typical process must offer a given set of input/output services to its communication environment, irrespective of its internal state. We show that under such outward-facing interactions, the correspondence is exact: a local mu-calculus property holds globally if, and only if, it holds locally.

We also detail the implications for entire families of networks that are defined by 'symmetry patterns.' For instance, a network family with a transitive global symmetry group can be analyzed by examining a single representative node. Such dramatic reductions in complexity are generally not possible for non-local properties.

None of the symmetry reduction results rely in any essential manner on the processes being finite-state. To summarize the main results:

– The local state spaces of balanced processes (the spaces incorporate interference from neighbors) are bisimilar. Hence, it suffices to model-check properties on representative processes of the balance equivalence classes,
– The local state space simulates the global space up to stuttering. Thus, a universal local mu-calculus property holds on the global space if it holds on a representative local space,
– With 'outward-facing' interaction, the local and global spaces are stuttering-bisimilar. A local mu-calculus property holds on the global space if, and only if, it holds on a representative local space.

We also explore the implications of these results and, in particular, show that in several settings, local symmetries can be determined easily from process syntax. We show that for isomorphic 'normal' processes operating in a network whose communication graph has at least transitive symmetry, a balance relation with a single representative process can be generated from the syntactic description of the network. In another direction, we show that for networks formed from 'building block' patterns, the pattern instances serve as balance representatives. These direct, syntactic constructions avoid having to build global symmetry reduced structures, can lead to exponential reductions in the cost of model checking, and apply to many networks where global symmetry reduction techniques are ineffective. Moreover, entire network families can be model-checked via the analysis of a small number of representative processes, so that the savings in the cost of analysis are unbounded.

## 2    Preliminaries

**Processes and Networks: Syntax.** A *network* is a directed graph, defined by a set of *nodes*, $N$, a set of *edges*, $E$, and two connection relations: $Out \subseteq N \times E$ and $In \subseteq N \times E$. Connections are directed from node $n$ to the edges in $Out(n)$, and directed inwards from the edges in $In(n)$ to $n$. Nodes $m$ and $n$ are neighbors, denoted $nbr(n,m)$, if they have a common connected edge. Node $m$ *points to* node $n$ if there is an edge $e$ in $Out(m) \cap In(n)$.

A *process* is defined by a tuple $(V, I, T)$, where $V$ is a set of variables which defines its local state space; $I(V)$ is a Boolean predicate defining the initial states; and $T(V, V')$ is a Boolean predicate defining the state transitions, using a copy $V'$ to denote the next state. Variables are partitioned into *internal* and *external* variables. External variables are labeled as *read*, or *write*, or both. The transition relation is required to preserve the value of read-only variables and its enabledness cannot depend on the values of write-only variables.

A *process network* $P$ is defined by a network graph, a set of processes, and an assignment, $\xi$. Every node $n$ is assigned a process $\xi(n)$, which we denote for convenience by $P_n = (V_n, I_n, T_n)$. Each edge $e$ is assigned a variable $\xi(e)$ in $V = (\bigcup n : V_n)$. The assignment $\xi$ must assign $In(n)$ to the read variables in $V_n$, $Out(n)$ to the write variables of $V_n$, and the internal variables of $V_n$ to no network edge. The *shared* variables of processes $P_m$ and $P_n$ are those assigned to common connected edges of $m$ and $n$.

**Processes and Networks: Semantics.** Semantically, the behavior of a process network $P$ is defined as the process $P = (I, V, T)$, where $V = (\bigcup n : V_n), I = (\bigwedge n : I_n)$, and $T = (\bigvee n : T_n \wedge \mathsf{unchanged}(V \setminus V_n))$. This defines an interleaving semantics, with $\mathsf{unchanged}(W)$ denoting that the values of variables in $W$ are unchanged.

A *global* state is a function mapping variables in $V$ to values in their domain. A *local* state of $P_n$ is a function mapping the variables in $V_n$ to values in their domain. An *internal* state of $P_n$ is a function mapping the internal variables of $P_n$ to values in their domains.

For neighbors $m, n$, a *joint state* is a pair $x = (x_m, x_n)$, where $x_m$ and $x_n$ are local states of processes $P_m$ and $P_n$, respectively, such that $x_m$ and $x_n$ have the same value for all shared variables. The transition relation $T_n$ is extended to joint states as $T_n(x, y)$, which holds iff $T_n(x_n, y_n)$ holds and the values of variables in $P_m$ that are not shared with $P_n$ are unchanged.

**Invariants: Global and Compositional.** Invariance is central to reasoning about dynamic system behavior. For a process network $P$ as defined above, a *global assertion*, $\theta$, is a set of global states of $P$. It is an *inductive invariant* for $P$ if all initial states are in $\theta$, i.e., $[I(x) \to \theta(x)]$, and $\theta$ is closed under transitions, i.e., $[\theta(x) \land T(x, y) \to \theta(y)]$.[1]

In place of a single invariance assertion, compositional reasoning postulates a set of *local assertions*, $\{\theta_n\}$, where $\theta_n$ is a set of local states of $P_n$, for each $n$. This set is a *compositional inductive invariant* if, for all $n$:

**(Init)** The initial states of $P_n$ are included in $\theta_n$. That is, $[I_n(x_n) \to \theta_n(x_n)]$
**(Step)** Transitions of $P_n$ preserve $\theta_n$. That is, $[\theta_n(x_n) \land T_n(x_n, y_n) \to \theta_n(y_n)]$
**(Non-Interference)** Assertion $\theta_n$ is preserved by transitions of neighbors $P_m$, from every joint state satisfying both $\theta_m$ and $\theta_n$. I.e., For all $m$ such that $nbr(n, m)$ and all joint states $x = (x_n, x_m), y = (y_n, y_m) : [\theta_n(x_n) \land \theta_m(x_m) \land T_m(x, y) \to \theta_n(y_n)]$

These constraints are in a simultaneous pre-fixpoint form over $\{\theta_n\}$. The least fixpoint is the strongest compositional invariant. For finite-state processes, this computation is polynomial-time in the size of the local state spaces.

**Theorem 1** [17]. *If $\{\theta_n\}$ is a compositional inductive invariant then $\bigwedge_i \theta_i$ is a global inductive invariant.*

**Symmetry Between Neighborhoods.** A neighborhood symmetry between nodes $m$ and $n$ is witnessed by a bijection, $\beta$, which maps edges in $In(m)$ to those in $In(n)$ and edges in $Out(m)$ to those in $Out(n)$; we call $(m, \beta, n)$ a similarity. The set of similarities $(m, \beta, n)$ is a groupoid[2].

A *balance* relation ([17], *c.f.* [11]) links symmetries throughout a network: balanced nodes $m, n$ have isomorphic neighborhoods, nodes connected to corresponding edges of $m, n$ are themselves balanced, and so on. Formally, a balance relation, $B$, is a set of triples $(m, \beta, n)$, such that $(m, \beta, n)$ is a similarity; $(n, \beta^{-1}, m)$ is in $B$; and for any node $k$ that points to $m$, there is a node $l$ which points to $n$ and a bijection $\gamma$ such that $(k, \gamma, l)$ is in $B$, and $\gamma(e) = \beta(e)$ for every edge $e$ that is connected to both $m$ and $k$.

The structure of this condition is similar to that of bisimulation (it is coinductive); thus, there is a greatest fixpoint, which is the largest balance relation. Nodes $m, n$ are *balanced* if $(m, \beta, n)$ is in the largest balance relation for some $\beta$.

---

[1] The notation, $[\varphi]$, from Dijkstra and Scholten [7], means that $\varphi$ is valid.
[2] I.e., $(n, \iota, n)$ is a similarity for the identity map $\iota$; if $(m, \beta, n)$ is a similarity, so is $(n, \beta^{-1}, m)$; and if $(m, \beta, q)$ and $(q, \gamma, n)$ are similarities, so is $(m, (\gamma\beta), n)$.

A process network $P$ *respects* balance relation $B$ if balanced nodes are assigned processes with isomorphic initial states and transition relations: i.e., for all $(m, \beta, n) \in B$, it is the case that $[I_n(\beta(s)) \equiv I_m(s)]$ for all $s$, and $[T_n(\beta(s), \beta(t)) \equiv T_m(s, t)]$ for all $s, t$. Similarly, we say that local assertions $\{\phi_i\}$ respect $B$ if $[\phi_n(\beta(s)) \equiv \phi_m(s)]$ for all $(m, \beta, n) \in B$. We abbreviate these conditions as $[I_n \equiv \beta(I_m)], [T_n \equiv \beta(T_m)]$ and $[\phi_n \equiv \beta(\phi_m)]$, respectively. Here, $\beta$ is overloaded to permute local states of $P_m$. For local state $s$ of node $m$, the local state $\beta(s)$ at node $n$ is defined as follows: the internal states of $m$ in $s$ and $n$ in $\beta(s)$ are identical and, for every edge $e$ connected to $m$, the value on $e$ in $s$ is identical to the value of $\beta(e)$ in $\beta(s)$. A key result is that balanced nodes have isomorphic compositional invariants.

**Theorem 2** ([17]). *If a process network respects balance relation $B$, its strongest compositional invariant also respects $B$.*

This theorem implies that it suffices to compute the strongest compositional invariant only for representative nodes[3], as the invariants for all other nodes are isomorphic to those of their representatives.

## 3   The Local Mu-Calculus

Intuitively, a local property is one that refers to the local state of a node, e.g., "the process at node $n$ is in its critical section", or "the philosopher at node $n$ holds all adjacent forks". We are interested in establishing a local property $f(n)$, parameterized by node $n$, and so isomorphic between nodes, for *all* nodes of a process network. We represent such a property by a mu-calculus formula. This has two interpretations: one in the global state space, the other in a compositionally constructed local state space. Their connections are discussed in the next section.

### 3.1   Syntax

The local mu-calculus syntax and semantics is largely identical to that of the standard $mu$-calculus [15]. The only difference is the use of the E[ U ] operator in place of EX, this is given a stuttering-insensitive semantics.

Let $\Sigma$ be a set of atomic propositions, $\Gamma$ be a set of propositional variables, and $\Delta$ a set of transition labels; these sets are mutually disjoint. Local mu-calculus formulas are defined by the following grammar. A formula is one of

– An atomic proposition from $\Sigma$,
– A propositional variable from $\Gamma$,
– $\neg \varphi$, for a formula $\varphi$,

---

[3] A balance relation $B$ induces the equivalence relation $m \simeq_B n$ if $(m, \beta, n) \in B$ for some $\beta$. The compositional fixpoint is calculated for a representative of each class of $\simeq_B$. In the fixpoint calculation, the assertion $\theta_n$ is replaced by $\gamma(\theta_r)$, where $r$ is the representative for $n$, and $\gamma$ is a chosen isomorphism such that $(r, \gamma, n)$ is in $B$.

- $\varphi \wedge \psi$, the conjunction of formulae $\varphi$ and $\psi$,
- $\mathsf{E}[\varphi \, \mathsf{U}_a \, \psi]$, where $\varphi, \psi$ are formulas, and $a$ is a transition label from $\Delta$,
- $\mu Z.\varphi(Z)$, where $\varphi(Z)$ is a formula syntactically monotone in $Z$ (i.e., all occurrences of $Z$ fall under an even number of negations).

Operators $\mathsf{A}[\varphi \, \mathsf{W}_a \, \psi] = \neg \mathsf{E}[\neg \varphi \, \mathsf{U}_a \, \neg \psi]$ and $\nu Z.\varphi(Z) = \neg \mu Z.(\neg \varphi(\neg Z))$ are the negation duals of $\mathsf{E}[\, \mathsf{U} \,]$ and $\mu$, respectively, with Boolean operations $\vee$ and $\rightarrow$ defined as usual.

### 3.2  Semantics

A state space has the form $(S, S_0, R, L)$, where $S$ is a set of states, $S_0$ is the set of initial states, $R \subseteq S \times \Delta \cup \{\tau\} \times S$ is a left-total transition relation, and $L : S \rightarrow 2^\Sigma$ labels states with atomic propositions. A path is a sequence $s_0, a_0, s_1, a_1, \ldots$ such that $(s_i, a_i, s_{i+1}) \in R$ for all $i$, where the sub-sequence $a_0, a_1, \ldots$ is the label sequence of the path.

The state set $S$ generates a complete lattice of all subsets of $S$, ordered by set inclusion. A functional $\Pi : 2^S \rightarrow 2^S$ is monotone if for all $A, B$ such that $A \subseteq B$ it is the case that $\Pi(A) \subseteq \Pi(B)$. By the Knaster-Tarski theorem, every monotone functional has a least and a greatest fixpoint. Consider a formula $\varphi(Z_1, \ldots, Z_d)$ with free variables $Z_1, \ldots, Z_d$. Given an assignment $\lambda$ mapping each free variable to a subset of $S$, the interpretation of $\varphi$ under $\lambda$ is defined inductively as follows. We write $M, s \models \varphi$ to mean that state $s$ in space $M$ satisfies a closed formula $\varphi$, i.e., $s$ is in $\mathsf{interp}(\varphi, \epsilon)$ where $\epsilon$ is the empty interpretation.

- $\mathsf{interp}(p, \lambda) = \{s \in S \mid p \in L(s)\}$, for proposition $p \in \Sigma$,
- $\mathsf{interp}(Z, \lambda) = \lambda(Z)$,
- $\mathsf{interp}(\varphi \wedge \psi, \lambda) = \mathsf{interp}(\varphi, \lambda) \cap \mathsf{interp}(\psi, \lambda)$,
- $\mathsf{interp}(\neg \varphi, \lambda) = S \setminus \mathsf{interp}(\varphi, \lambda)$,
- State $s$ is in $\mathsf{interp}(\mathsf{E}[\varphi \, \mathsf{U}_a \, \psi], \lambda)$ if, and only if, there is a finite path $\pi$ from $s$ to state $t$ with label sequence $\tau^*; a$, where $t$ is in $\mathsf{interp}(\psi, \lambda)$ and every other state $s'$ on $\pi$ is in $\mathsf{interp}(\varphi, \lambda)$. Informally, $\varphi$ holds until the first $a$-action, after which $\psi$ is true,
- $\mathsf{interp}(\mu Z.\varphi(Z), \lambda)$ is the least fixpoint of functional $\Pi(X) = \mathsf{interp}(\varphi(Z), \lambda')$ where $\lambda'$ extends $\lambda$ with the assignment of $X$ to $Z$.

### 3.3  Local and Global Interpretations

Let $\theta$ be a compositional invariant respecting a balance relation $B$. For any node $n$ of the network, define $H_n^\theta$ as the following transition system:

- The states are the local states of $P_n$ that satisfy $\theta_n$,
- A transition $(s, s')$ is either
  - A transition (labeled with $n$) by $P_n$ from state $s$, or
  - An interference transition (labeled with $m$) by a neighbor $P_m$ from a joint state $(s, u)$ where $\theta_n(s)$ and $\theta_m(u)$ hold, to a joint state $(s', u')$.
  
  By the properties of a compositional invariant, $s'$ is in $\theta_n$ in both cases.

The only missing ingredient is a labeling of the states with atomic propositions. Given such a labeling, $L$, a closed formula evaluates to a set of local states.

The global transition system $G$ defines the semantics of the process network. For a given $n$, let $G_n$ be $G$ with transitions by $P_n$ labeled with $n$, transitions by neighbors $m$ of $n$ labeled with $m$, and all other transitions (which cannot change the local state of $P_n$) labeled with $\tau$. A local labeling $L$ of $P_n$ is extended to $G_n$ by labeling a global state $s$ with proposition $p$ if $p$ labels the local state of $P_n$ in $s$. Formulas local to node $n$ are evaluated over $G_n$. A closed formula evaluates to a set of global states.

### 3.4 Simulation and Bisimulation

For processes without $\tau$ actions, a simulation relation $\alpha$ from process $P$ to process $Q$ is a relation from the state space of $P$ to that of $Q$, satisfying:

– Every initial state of $P$ is related to an initial state of $Q$ by $\alpha$, and
– If $s\alpha t$ holds, then $s$ and $t$ satisfy the same atomic propositions, and
– If $s\alpha t$ holds and $s'$ is a successor state of $s$ in $P$, there is a successor state $t'$ of $t$ in $Q$ such that $s'\alpha t'$ holds.

If a simulation relation exists from $P$ to $Q$, we say that $Q$ simulates $P$. It is well known that if $Q$ simulates $P$, then any standard universal mu-calculus formula that holds for all initial states of $Q$ also holds for all initial states of $P$. A universal local mu-calculus formula is one where its negation normal form does not contain $\mathsf{E}[\,\mathsf{U}\,]$. Relation $\alpha$ is a bisimulation from $P$ to $Q$ if $\alpha$ is a simulation from $P$ to $Q$ and $\alpha^{-1}$ is a simulation from $Q$ to $P$. It is well known that bisimilar processes satisfy the same standard mu-calculus properties.

For processes with $\tau$ transitions, one can relax the third condition to allow the possibility of stuttering (cf. [4]): if $s\alpha t$ holds, then for any state $s'$ reachable from $s$ by a finite path $\pi$ with label sequence $\tau^*; a$ (for a non-$\tau$ letter $a$), there is a state $t'$ reachable from $t$ by a finite path $\delta$ labeled $\tau^*; a$ such that $s'$ and $t'$ are related by $\alpha$, and every other pair of states $u$ on $\pi$ and $v$ on $\delta$ is related by $\alpha$. Relation $\alpha$ is a stuttering bisimulation if $\alpha$ and $\alpha^{-1}$ are stuttering simulations.

**Theorem 3.** *Stuttering simulation preserves universal local mu-calculus properties. Stuttering bisimulation preserves all local mu-calculus properties.*

## 4 Connecting Local Mu-Calculus Interpretations

We explore relationships between the local and global interpretation of formulas, and show the following:

– The local state spaces of balanced nodes are bisimilar. It follows from Theorem 3 that balanced nodes satisfy the same local mu-calculus formulas. From this result, to model check a property of the form $(\bigwedge i :: f(i))$, it suffices to check $f(i)$ for the representatives of the balance equivalence classes.

- The local state space of node $m$ stuttering-simulates the global state space up to the local state of $m$. It follows from Theorem 3 that a universal local mu-calculus formula on $m$ holds globally if it holds locally.
- If processes exhibit 'outward-facing' interaction, i.e., (roughly) the effect of interfering transitions is independent of the internal state of the interfering process, then the local and global state spaces are stuttering-bisimilar up to the local state of $m$. It follows that the two spaces satisfy precisely the same local mu-calculus formulas over $m$.

*Notation.* In the proofs below, for a local state $s$ of node $n$, the notation $s[n]$ refers to the internal state of $P_n$ in $s$, and for an edge $e$ that is connected to $n$, the notation $s[e]$ refers to the value in $s$ of the variable assigned to $e$.

### 4.1 Bisimilarity Between Local State Spaces

**Theorem 4.** *Let $B$ be a balance relation on a process network $P$, and $\theta$ a compositional invariant for the network. If $P$ and $\theta$ respect $B$, then for every $(m, \beta, n)$ in $B$, $H_m^\theta$ and $H_n^\theta$ are bisimilar up to $\beta$.*

**Proof:** The bisimulation relation $R$ relates a local state $s$ of node $m$ to a local state $t$ of node $n$ if $\beta(s) = t$. Before getting to the details of the proof, which is technical, we sketch the main reasoning. First, local transitions are easily matched by symmetry. For an interfering transition from a neighbor $k$ of $m$, by balance, there is a matching neighbor $l$ of $n$ with a symmetric interference transition. Crucially, the preservation of the compositional invariant under balance lets us transfer the joint state from which the interference transition occurs in $H_m^\theta$ to a joint state with a matching interference transition in $H_n^\theta$.

Suppose that $s, t$ are states of $m$ and $n$ in the local state spaces $H_m^\theta$ and $H_n^\theta$, respectively, such that $sRt$ holds, that is $\beta(s) = t$. By construction of $H_m^\theta$ and $H_n^\theta$, $\theta_m(s)$ and $\theta_n(t)$ hold.

Consider a step transition $T_m(s, s')$. Since $T_m$ and $T_n$ respect the balance relation, $B$, by the local symmetry between the transition relations, $T_n(\beta(s), \beta(s'))$ holds as well. Thus, for $t' = \beta(s')$, we have that there is a step transition $T_n(t, t')$ such that $s'Rt'$. By construction, $s'$ and $t'$ are successors of $s$ and $t$, respectively, in the local state spaces.

Now consider an interference transition in $H_m^\theta$ from a joint state $(s, u)$ where $u$ is a local state of a neighbor $k$ of $m$. The transition $T_k(u, u')$ creates a joint state $(s', u')$. From the definition of balance, there is a neighbor $l$ of $n$ such that for some $\gamma$, we have $(k, \gamma, l)$ in the balance relation. As $\theta$ respects $B$ by assumption, we have that $\theta_l = \gamma(\theta_k)$. As $\theta_k(u)$ holds by the definition of the interference transition, the state $v = \gamma(u)$ is in $\theta_l$. We claim that there is a matching transition from the joint state $(t, v)$.

First, we show that the pair $(t, v)$ forms a joint state. Consider any edge $f$ that is shared between $n$ and $l$. By balance, shared edges are mapped identically by $\beta$ and $\gamma$; hence, $e = \beta^{-1}(f) = \gamma^{-1}(f)$ is shared by $m$ and $k$. By the definition of $t = \beta(s)$ and $v = \gamma(u)$, we have that $t[f] = s[e]$ and $v[f] = u[e]$. As $(s, u)$ is a joint state, we have $s[e] = u[e]$; hence, $t[f] = v[f]$. As $f$ was chosen arbitrarily, it follows that $t$ and $v$ agree on the values of all shared edges, so $(t, v)$ is a joint state. Moreover, the state $t$ is in $\theta_n$ by assumption, and $v$ is in $\theta_l$ by construction.

By the similarity between $P_k$ and $P_l$, there is a transition $T_l(\gamma(u), \gamma(u'))$; letting $v' = \gamma(u')$, this can be expressed as $T_l(v, v')$. That induces an interference transition in $H_n^\theta$ from the joint state $(t, v)$ to a joint state $(t', v')$.

Finally, we show that $t' = \beta(s')$. Let $e$ be an edge connected to node $m$ and let $f = \beta(e)$. Note that $f$ is shared between $n$ and $l$ if, and only if, $e$ is shared between $m$ and $k$. Now if $f$ is not shared between $n$ and $l$, then $t'[f] = t[f]$ by definition of interference; $t[f] = s[e]$ as $t = \beta(s)$; and $s'[e] = s[e]$ by definition of interference. By transitivity, $t'[f] = s'[e]$, as required. If $f$ is a shared edge, then $t'[f] = v'[f]$ by joint state; $v'[f] = u'[e]$ as $v' = \gamma(u')$; and $u'[e] = s'[e]$ by joint state. By transitivity, $t'[f] = s'[e]$. The internal states of $t, t'$ and $s, s'$ are (respectively) identical, as they are unaffected by interference. Hence, $t' = \beta(s')$.

The proof so far shows that $R$ is a simulation if $(m, \beta, n)$ is in the balance relation. From the same argument applied to $(n, \beta^{-1}, m)$, which must also be in the balance relation, the inverse of $R$ is also a simulation. Hence, $R$ is a bisimulation between $H_m^\theta$ and $H_n^\theta$. **EndProof.**

We say that per-process propositional labelings *respect* balance if for every $(m, \beta, n)$ in the balance relation, every atomic proposition $p$, and every local state $s$: $[p \in L_n(\beta(s)) \equiv p \in L_m(s)]$. From Theorems 3 and 4, we obtain:

**Corollary 1.** *Let $f(i)$ be a local mu-calculus formula parameterized by $i$. If the compositional invariant $\theta$ and the interpretation of the atomic propositions in $f$ respect balance relation $B$, then for any $(m, \beta, n)$ in $B$ and any local state $s$: $H_m^\theta, s \models f(m)$ if, and only if, $H_n^\theta, \beta(s) \models f(n)$.*

### 4.2  Local-Global Simulation

From the point of view of a process $P_m$, a transition in the global state space is either a transition of $P_m$, or an interference transition by one of the neighbors of $m$, or a transition by a "far away" process that has no immediate effect on the local space of $m$. Thus, global transitions can be simulated by step or interference transitions in the local space, with far-away transitions exhibiting stuttering. The converse need not be true, as interference transitions appear in the local space without the constraining context of the entire global state.

**Theorem 5.** *Let the scheduling of transitions in the global system be unconditionally fair. For every $m$ and any compositional inductive invariant $\theta$, $H_m^\theta$ simulates the global transition system $G_m$ up to stuttering.*

**Proof:** For a global state $s$, let $s[m]$ refer to the local state of node $m$ in $s$. Define the relation $R$ from global states to those of $H_m^\theta$ by $(s, t) \in R$ iff $\theta(s)$ and $s[m] = t$. We show that $R$ is a simulation, up to stuttering. The proof is by cases on the kinds of transitions from global state $s$ to a successor state, $s'$. As $\theta$ is a global *inductive* invariant by Theorem 1, it is the case that $\theta(s')$ holds.

Suppose the transition is by process $m$. Thus, $T_m(s[m], s'[m])$ should hold. As $\theta_m(s[m])$ holds, this transition is in the local state space as well. Letting $t' = s'[m]$, we have $s'Rt'$.

Suppose the transition is by a neighbor $k$ of $m$, so that $T_k(s[k], s'[k])$ holds, and for all edges $e$ that are not connected to $k$, $s'[e] = s[e]$. By definition, $\theta_m(s[m])$ and $\theta_k(s[k])$ hold, so this is a valid interference transition in the local state space $H_m^\theta$. Denoting $s[k]$ by $u$, this can be re-expressed as a joint transition from state $(t, u)$ to $(t', u')$, where $u' = s'[k]$. Consider an edge $e$ that is connected to $m$ but not to $k$. Then $t'[e] =$ (by non-adjacency) $t[e] =$ (by $R$) $s[m][e] =$ (by non-adjacency) $s'[m][e]$. Now consider an edge $e$ that is shared by nodes $m$ and $k$; then $t'[e] =$ (by shared edge) $u'[e] =$ (by definition) $s'[k][e] =$ (by shared edge) $s'[m][e]$. The internal state of $m$ is unchanged on either transition. Thus, $t' = s'[m]$, so that $s'Rt'$, as desired.

Finally, suppose the transition is by a process that is not a neighbor of $m$. Then $s'[m] = s[m]$, so that $s'Rt$ holds. This is the stuttering step. As transitions are scheduled in an unconditionally fair manner, on any infinite computation from $s$, process $m$ or one of its neighbors must eventually make a move. Hence, all stuttering is bounded. This establishes (fair) stuttering simulation between the two spaces. **EndProof.**

From the preservation of universal local mu-calculus properties under stuttering simulation, we have:

**Corollary 2.** *If $f(m)$ is a universal local mu-calculus formula, then for any $t, s$ such that $s[m] = t$: $H_m^\theta, t \models f(m)$ implies that $G_m, s \models f(m)$ under fairness.*

### 4.3 Outward-Facing Interactions and Local-Global Bisimulation

The obstacle to establishing bisimilarity in the proof of Theorem 5 is that an interference transition from local state $t$ may not have a corresponding transition from a related global state $s$, as the internal state of the interfering neighbor in $s$ may be different from the internal state of the interfering neighbor of $t$. In some protocols, however, we see that interference depends only on the shared state. For instance, in a form of the dining philosophers' protocol where a process may give up a fork if it is not eating, the interference transition (passing a fork to a neighbor) is dependent only on possession of the fork. In this setting, one can indeed show that the two spaces are bisimilar.

We express the independence from internal state as a stuttering bisimulation within the interfering process. Define a relation $B_{m,n}$ on the local state space of $P_n$ by $(u, v) \in B_{m,n}$ if $u$ and $v$ are both in $\theta_n$, and $u[e] = v[e]$ for every edge $e$ shared between $m$ and $n$. We say that process $n$ is *outward-facing* in interactions with its neighbor $m$ if the relation $B_{m,n}$ is a stuttering bisimulation on $H_n^\theta$.

**Theorem 6.** *With outward-facing interaction, the local state space of process $m$ is stuttering bisimilar to the global state space in terms of the local state of $m$.*

**Proof:** Define the relation $R$ from global states to those of $H_m^\theta$ as in the proof of Theorem 5 by $(s, t) \in R$ iff $\theta(s)$ and $s[m] = t$.

Consider a transition from $t$ to $t'$. If the move is by process $m$, it is enabled in $s$ as well, and the resulting states are related by $R$. Now suppose the move is an interference transition by a neighbor, $n$. Hence there is some joint state $(t, u)$ of $(m, n)$ such that the move is by $n$ from $(t, u)$ to $(t', u')$. As $u \in \theta_n$ (by joint state) and $s[n] \in \theta_n$ (by definition of $R$), and the two are connected to the same local state of $m$, the pair $(s[n], u)$ is in $B_{m,n}$. As $B_{m,n}$ is a stuttering bisimulation, there is a sequence, say $\sigma$, of transitions by $P_n$ alone from $s[n]$ to a state $v'$ such that $(v', u') \in B_{m,n}$, and all intermediate states on $\sigma$ from $s[n]$ to $v'$ are related by $B_{m,n}$ to $u$. Hence, the value of the shared edges between $m$ and $n$ is unchanged on $\sigma$ until the final step, where it matches $u'$. Therefore, for the global computation induced by $\sigma$ from $s$, the final state $s'$ is such that $s'Rt'$, and for all intermediate global states $x$ on that path, $xRt$ holds. This shows that $R^{-1}$ is a stuttering simulation from the local to the global space. By Theorem 5, the relation $R$ is a simulation from the global to the local space. Hence, $R$ is a stuttering bisimulation between the spaces. **EndProof.**

**Corollary 3.** *With outward-facing interaction and unconditionally fair scheduling, the local state space of a process $m$ satisfies the same local mu-calculus properties as the global state space.*

## 5  Syntactic Determination of Local Symmetries

We show how to recognize local symmetry from syntactic structure. This also applies to network families, with corresponding unbounded savings in local verification. First, we use relations between structure and global symmetry, and between global and local symmetries. Next, we show how local symmetries may be directly derived if network families are induced by a finite set of tilings. We note that when local symmetry is derived syntactically, either through the use of normal process descriptions, or through building block tiles, the computation of the compositional invariant can be done symbolically, and in the case of tilings, directly on each tile, unlike the case of global symmetry reduction, where the symbolic (BDD-based) orbit relation is difficult to compute even for fully symmetric networks [5].

### 5.1  Program Symmetries

Let $P = ||_{i \in [0..k-1]} P_i, k \geq 1$ be a fixed network where each component $P_i$ is an implementation of a process template $W$. Network topology is restricted so that all edges are bidirectional and connect only two nodes. Each $P_m$ is described by a finite transition graph where if there is an arc from the internal node $g$ to the internal node $h$ then the arc is labeled by a guarded command $\rho \to A$. Transitions are given by $g : \rho \to A : h$ where $A$ is the local update function and $\rho$ is a predicate over the neighborhood of $P_m$. The action $A$ is given by a list of simultaneous updates to the shared variables, $v_1, \ldots, v_d$, where $v_i$ is the variable across the edge $(m, n_i)$.

We name the variables associated with a process, depending on the specific topology, the left variable, the right variable, the forward variable of $P_m$, etc. This modeling tactic is used (see [8]) to stipulate that the update functions for the variables be process-index independent.

Two transitions $g : \rho \to A : h$ and $g' : \rho' \to A' : h'$ are equivalent if $g = g', h = h', \rho$ is semantically equivalent to $\rho'$ and $A$ and $A'$ are semantically equivalent (*c.f.* [8]). Processes $P_m$ and $P_n$ are equivalent if there is a bijective mapping between equivalent transitions of $P_m$ and $P_n$. A permutation $\pi$ of process indices is an automorphism of $P$ if $P_m$ is equivalent to $P_{\pi(m)}$ for all $m \in [0..k-1]$.

As shown in [8] the global symmetries of the program $P$, essentially the permutations of $[0..k-1]$ that leave $P$ unchanged, are a subset of the global symmetries of the global state space $G$. From $P$, one defines an undirected graph, the *communication relation*, $CR$ [8]. The nodes of $CR$ are the nodes of $N$ of the topology $(N, E)$ and there is an edge from $m$ to $n$ in $CR$ iff the nodes are connected to a common edge.

$P$ is *normal* [8] if the transitions of $P$ are given in the following form:

$$g : (\wedge_{n \in CR(m)} \rho(m,n)) \to (\wedge_{n \in CR(m)} A(m,n)) : h$$

where each $\rho(m,n)$ is a boolean expression over the internal state of $P_m$ and the neighborhood variables of $P_m$, or equality tests between the variables local to the neighborhood of $P_m$, and the assignments of $A(m,n)$ are concurrent assignments to the neighborhood variables of $P_m$, where variable values may be swapped with each other or assigned constant values. When $P$ is a normal process network [8] showed that global symmetries of $CR$ are symmetries of $P$ and are automorphisms of $G$.

This setting substantially simplifies the application of local symmetry. First, the balance relation can be "read off" directly from the relation $CR$, as by results in [17], the global symmetries of $CR$ define a balance relation over $(N, E)$, which includes $(m, \beta, n)$ if there is a symmetry $\pi$ of $CR$ such that $\pi(m) = n$. Secondly, if $CR$ induces a transitive symmetry group, then local symmetry reduction reduces to analysis of a single representative process and its neighborhood. This may result in an exponential reduction in the cost of model checking, compared with an analysis of the entire state space. (The global symmetry used in [8] provides an exponential reduction only when $CR$ is fully symmetric.) The check is in general over-approximate (cf. Corollary 2) but is exact under outward-facing interaction. In the parametric setting, the reduction is unbounded.

### 5.2 Tilings

Rings, tori, and other 'regular' network patterns have considerable local symmetry but little global symmetry. Here we show how to enforce local symmetry across network families by generating them from a finite set of *tiles*. The tiles directly induce local symmetries and balance.

Consider a fixed, finite set of process types where each process type has a fixed, finite set of edge directions, which are given unique names. The initial condition and the transition relation of a process type may refer to the values on edges in the given direction. Each type is associated with a tile describing a fixed neighborhood pattern around a node of that type. The pattern specifies for each edge connected to the central node its direction from the center and the type and direction of the other process connected to it. The tiles induce a family of networks, typically of unbounded size, as follows. A network is in the family if (1) each node is assigned an instance of a process type, and (2) the neighborhood of a node matches the tile for that node type. For instance, a tile for a torus shape would have 4 neighbors, labeled north, south, east and west.

A network family constructed in this manner has an induced balance relation, $B$, defined as follows. Let $m, n$ be nodes of a network in the family. Let $(m, \beta, n)$ belong to $B$ if (a) both nodes are instances of the same type and (b) $\beta$ is the mapping which, for each direction $a$, relates the edge reachable in direction $a$ from $m$ to the edge reachable in the same direction from $n$. (E.g., it maps the north edge of $m$ to the north edge of $n$.)

**Theorem 7.** *$B$ is a balance relation for the induced family, with finitely many equivalence classes.*

**Proof:** We show that $B$ is a balance relation, and that it is respected by the process assignment. The mapping $\beta$ is an isomorphism of the edges connected to $m$ and $n$, as both have the same type. Moreover, as their initial conditions and transition relations are derived from those of the type and are independent of node identities, they are isomorphic up to $\beta$.

We now establish that $B$ meets the balance relation. Consider a direction $a$. Let $m'$ ($n'$) be the node connected to $m$ ($n$) in that direction. As $m$ and $n$ have the same tiling pattern, $m'$ and $n'$ have the same type, so the tuple $(m', \gamma, n')$ is in $B$, for the isomorphism $\gamma$ between the edges of $m'$ and $n'$ as given in the definition of $B$. Consider the edge $e$ reached from $m$ in direction $a$, and let $b$ be the direction that this edge is reached from $m'$. Let $f$ be the edge in direction $a$ from $n$. As $m$ and $n$ follow the same tiling pattern, $f$ must be reached from direction $b$ from $n'$. Therefore, $\beta$ and $\gamma$ agree on this edge. As the edge was chosen arbitrarily, this establishes the balance condition. The number of equivalence classes induced by the greatest balance relation is, then, at most the number of tiles, which equals the number of process types. **EndProof.**

Theorem 7 implies that the compositional analysis of all instances of the network family can be reduced to the analysis of a finite set of representatives. This contrasts with global symmetry reduction for network families, where parameterized collapse is not as simple, nor as general. Moreover, the required representatives are just the tiles. The easy syntactic symmetry reduction contrasts with the difficulty of computing global symmetry groups for network families.

## 6   Applications

**Example 1.** Consider a non-deterministic token-ring system $P = \|_i P_i$. The internal states of $P_i$ range over $\{\texttt{T}, \texttt{H}, \texttt{E}\}$ with shared variables $x_i$ and $x_{i+1}$ ranging over $\{\bot, tok\}$. Initially, each process is in internal state $T$ and either owns 0 tokens or owns 1 token. The initial condition specifies that a single process owns the token. Processes cycle through states in the order $T, H$ and $E$. A process in $H$ can move to $E$ only if it owns the token. When exiting $E$ the process puts the token on its right and enters $T$. If a process is in $T$ and has the token, then it either enters $H$ or passes the token to the right. It can be shown that the process interactions are outward-facing. Verification of the mutual exclusion property *for all i*: $\mathsf{AG}(E_i \to (x_i = tok))$ can then be performed on a model with 3 processes that suffices to see all reachable local states.

In addition, a liveness property, *for all i* : $\mathsf{AG}(H_i \to \mathsf{AF}E_i)$, can also be verified using a combination of local arguments. The proof is constructed as follows: first, show that the system satisfies the invariant that there is exactly 1 token in the system. Then show every process that has the token eventually passes the token to the neighbor on the right. Using the global system fairness assumption that each process executes infinitely often we can chain these proofs together to conclude that for any particular process $P_n$: $\mathsf{AG}(H_n \to \mathsf{AF}E_n)$ holds which by local symmetry implies: *for all i* : $\mathsf{AG}(H_i \to \mathsf{AF}E_i)$.

**Example 2.** Interestingly, the results about a single token ring network can be extended to a ring with 2 tokens. However, the minimal model requires 4 processes. Similar reasoning holds for 3 tokens and we hypothesize can be generalized to any fixed number of tokens. A related example is a ring with 2 types of processes, one labeled *red* and one labeled *black*. For rings with even numbers of processes, half of them *red* and half of them *black*, there are 2 equivalence classes. Local symmetry reduction can be used to verify behavior of the two equivalence classes for any even number of processes, though the networks have little global symmetry and do not have transitive symmetry.

**Example 3.** Several works including [3,9,10,14] have considered using counting arguments as a way of implementing full symmetry reduction. Given an $n$ process system, with isomorphic processes having local state spaces of size $m$, and full global symmetry on $[1..n]$ the idea is to replace the global symmetry-reduced model with a set of $m$ counters, where the counter values record the number of components in each of the different local states. A combinatorial argument [22] shows that the number of combinations of $n$ isomorphic process each with $m$ local states, is $(m + n - 1)!/(n!(m - 1)!)$. If $n > 2m$, this is more than $2^m$. On the other hand, if each component has $b$ neighbors, the local representative (full global symmetry implies a single balance class) has a local state space of size approximately $m^b$. Over a parametric analysis $m^b$ is a constant and $b$, the number of neighbors, is likely to be small in comparison with $m$.

## 7   Discussion and Related Work

We studied the relationship between the satisfaction of temporal properties on the global state space of a process network and on individual local state spaces. We show that "balanced" processes have bisimilar local spaces and therefore satisfy the same local mu-calculus formulas. Hence, for a local formula $f(n)$ that is universal in nature, the satisfaction of $f(n)$ on the local space of node $n$ implies that $f(n)$ holds of the global state space. Thus, if universal formulas $\{f(n)\}$ hold for all nodes $n$, then $(\bigwedge i : f(i))$ holds for the global state space. This provides an approximate way to establish quantified mu-calculus properties. Moreover, as balanced nodes satisfy the same formulas, it is only necessary to model-check representatives of the balance equivalence relation. For a fixed process network, the restriction to local state spaces can result in exponential savings (in the number of nodes), and the further restriction to representative spaces results in a further linear cost saving. More dramatically, we show that network families constructed from building-block "tiles" have a finite set of representative nodes, so the cost saving is unbounded for parametric analysis. When network processes communicate with their neighbors in an outward-facing manner, these results carry over to the entire local mu-calculus, not just to universal properties.

The results build on our earlier work on balance relations and local symmetry [17,18,20]. That work focused on compositional invariants [21] the central result being that the strongest compositional invariants for balanced nodes are isomorphic. The current paper shows that the isomorphism applies to all local mu-calculus properties. The local state spaces on which the mu-calculus properties are evaluated are built using compositional invariants. An elegant methodology using 3-valued logic to compositionally verify mu-calculus properties is developed in [23]; however, it applies to pairs of processes, and thus does not consider symmetries in larger networks. The definition of network families through tilings has similarities to the network grammars used in [24,26]; however, the verification techniques are different.

The framework of this paper considers the neighborhood of a single node. Compositional invariants have been generalized to apply to groups of processes, to accommodate properties stated over all pairs $i, j$, or over all neighbors $i, j$; see for example [1,6,12,13,16]. Construction of a comprehensive theory of neighborhood symmetry for groups of processes is still an open question.

Global symmetry reduction, developed in [5,8,14], is based on a beautiful mathematical theory of automorphisms in graphs. However, in practice, symmetry reduction runs into difficulties, usually because there is not enough global symmetry in a process network, but also because for even highly symmetric networks, symbolic manipulation of symmetry reduced structures is difficult. In fact [5] shows that any BDD-based representation of the global symmetry group for any network with only transitive symmetry would likely incur a prohibitive cost. By focusing on local similarities, a strict generalization of global symmetries [17,20], we can avoid these problems and obtain exponential improvements. The theory of local symmetries is based on network groupoids, and we note that any network automorphism group induces a balance relation.

We also consider parameterized verification. For network families built from building-block tiles, there is a finite set of representative neighborhoods, and it suffices to prove a parameterized local mu-calculus property for each of those representatives to show that it holds for the entire family. This is an approximate method for parameterized verification. In prior work [20], we had introduced the local PCMCP (parameterized compositional model-checking) question as a decision problem that is, in many cases, more tractable than the global PMCP (parameterized model-checking) problem. Deciding PCMCP for local mu-calculus properties is a challenging open question.

**Acknowledgements.** Kedar Namjoshi was supported, in part, by grant CCF-1563393 from the National Science Foundation. Richard Trefler was supported, in part, by an Individual Discovery Grant from the Natural Sciences and Engineering Research Council of Canada. Both authors thank E. Allen Emerson for inspiring discussions on the topic.